\pdftrailerid{}

\documentclass[sigconf,nonacm]{acmart}
\usepackage{float}
\usepackage{array}
\usepackage{booktabs}
\usepackage{colortbl}
\usepackage{enumitem}
\usepackage{graphicx}
\usepackage{makecell}
\usepackage{multirow}
\usepackage{pifont}
\usepackage{tabularx}
\usepackage{tikz}
\usepackage{xspace}
\usetikzlibrary{arrows.meta,calc,decorations.pathreplacing,positioning,shapes.geometric}
\definecolor{PaperBlue}{HTML}{3B5B92}
\definecolor{PaperTeal}{HTML}{2A8C82}
\definecolor{PaperAmber}{HTML}{D18A20}
\definecolor{PaperVermilion}{HTML}{C4543C}
\definecolor{PaperPurple}{HTML}{7561A8}
\definecolor{PaperOlive}{HTML}{73823D}
\definecolor{PaperInk}{HTML}{2F3437}
\definecolor{PaperGray}{HTML}{6B7280}
\definecolor{PaperLight}{HTML}{EDF2F6}

\newcommand{\systemname}{PackServe\xspace}
\newcommand{\papertitle}{PackServe: SLO-Aware Request Scheduling for Agentic LLM Serving at Scale}
\newcommand{\papertitleplain}{PackServe: SLO-Aware Request Scheduling for Agentic LLM Serving at Scale}
\newcommand{\papershorttitle}{PackServe: SLO-Aware Request Scheduling}

\newcommand{\systemnamefootnote}{}
\newcommand{\systemnameplain}{PackServe}

\graphicspath{{figures/}}
\newif\ifearlyeoverview
\newif\iflateeplots

\newcolumntype{L}[1]{>{\raggedright\arraybackslash}m{#1}}
\newcolumntype{C}{>{\centering\arraybackslash}X}
\floatstyle{ruled}
\newfloat{algorithm}{tbp}{loa}
\floatname{algorithm}{Algorithm}
\newcounter{algorithmline}[algorithm]
\newcommand{\algline}[1]{%
  \refstepcounter{algorithmline}\label{#1}\thealgorithmline:}
\newcommand{\yes}{\ding{51}}
\newcommand{\no}{\ding{55}}

\earlyeoverviewtrue
\lateeplotstrue
\makeatletter
\renewcommand{\@subsubsecfont}{\sffamily\bfseries\upshape}
\renewcommand{\@parfont}{\bfseries\upshape}
\let\paper@startsection\@startsection
\def\@startsection#1#2#3{%
  \ifstrequal{#1}{paragraph}%
    {\paper@startsection{#1}{#2}{\z@}}%
    {\paper@startsection{#1}{#2}{#3}}}
\makeatother
\title[\papershorttitle]{\papertitle}
\acmDOI{}
\acmISBN{}
\hypersetup{hypertexnames=false,keeppdfinfo=true,
  pdftitle={\papertitleplain},pdfpublisher={Authors},
  pdfpublication={Research preprint},pdfpubtype={article}}
\author{Zhiyuan Tan}
\authornote{These authors contributed equally to this work.}
\affiliation{%
  \institution{The Chinese University of Hong Kong, Shenzhen}%
  \country{China}%
}
\email{zhiyuantan1@link.cuhk.edu.cn}
\author{Dejiang Zhu}
\authornotemark[1]
\affiliation{%
  \institution{Ant Group}%
  \country{China}%
}
\email{zhudejiang.pt@antgroup.com}
\author{Jingzhe Jiang}
\affiliation{%
  \institution{The Chinese University of Hong Kong, Shenzhen}%
  \country{China}%
}
\email{226085102@link.cuhk.edu.cn}
\author{Yihao Zheng}
\affiliation{%
  \institution{The Chinese University of Hong Kong, Shenzhen}%
  \country{China}%
}
\email{225040496@link.cuhk.edu.cn}
\author{Yang Tian}
\affiliation{%
  \institution{Ant Group}%
  \country{China}%
}
\email{lieyuan@antgroup.com}
\author{Tao Wang}
\authornote{Corresponding authors.}
\affiliation{%
  \institution{Ant Group}%
  \country{China}%
}
\email{junchen.wt@antgroup.com}
\author{Minchen Yu}
\authornotemark[2]
\affiliation{%
  \institution{The Chinese University of Hong Kong, Shenzhen}%
  \country{China}%
}
\email{yuminchen@cuhk.edu.cn}
\renewcommand{\shortauthors}{Tan et al.}

\begin{document}
\begin{abstract}

Request scheduling is a key challenge in large-scale clusters serving agentic large language model (LLM) workloads.
An effective scheduler must preserve key--value cache (KVC) reuse across long, shared prefixes, meet token-level latency service-level objectives (SLOs), and minimize GPU resource footprint.
Existing schedulers struggle to reconcile these requirements: request consolidation can sacrifice cache locality and increase prefill/decode interference, compromising both SLO attainment and resource efficiency.
We present \systemname, a scheduler designed to reduce resource costs while meeting latency SLOs for agentic LLM serving.
\systemname uses compact white-box models to predict latency under prefill/decode interference. 
Guided by these predictions, it packs requests onto fewer serving instances while preserving KVC reuse and SLO constraints, trading available latency headroom for improved per-instance throughput.
Evaluation on 64 NVIDIA H20 GPUs shows that \systemname uses up to 16.8\% and 24.6\% fewer GPU-hours than state-of-the-art schedulers under 30-ms and 50-ms TPOT targets, respectively, while meeting the target TPOT objectives.
\systemname has also been deployed in our production cluster comprising over 1000 GPUs, where it reduces the resource footprint by 34.7\% compared with the original production scheduler.

\end{abstract}

\maketitle
\hypersetup{pdfsubject={Research preprint}}
\section{Introduction}
\label{sec:introduction}

Recent LLMs~\cite{deepseekai2026deepseekv41flash,glm52026,kimik32026,gpt6astra2026,claudefable52026} demonstrate strong capabilities in agentic tasks involving coding and tool use.
Agentic applications such as Codex~\cite{openaiCodex2025},
Claude Code~\cite{anthropicClaudeCode2025}, and OpenClaw~\cite{openclaw2026}
are therefore becoming increasingly significant workloads for cloud-based large language model (LLM) serving.
They execute user-specified tasks (e.g., code generation) within \emph{sessions}, which comprise multi-step workflows that interleave LLM requests with tool execution~\cite{react2023,toolformer2023,sglang2024,autogen2024,sweagent2024}.
Within each session, tool results and execution history are incorporated into subsequent inputs, forming request chains with long, shared prefixes~\cite{copilottraces2026,tracelab2026,agentsysbench2026}.

To serve agentic workloads at scale, cloud providers deploy clusters of GPU-backed inference instances, with gateways scheduling each incoming request to an instance~\cite{lmetric2026} (see Fig.~\ref{fig:system-overview}).
Each request proceeds through two phases: \emph{prefill} processes the input and produces the first output token, while \emph{decode} generates subsequent tokens autoregressively~\cite{sarathiserve2024}.
We focus on deployments that colocate both phases on the same GPUs, a configuration widely used in production serving systems~\cite{lmetric2026}.
In this setting, request scheduling determines cache locality and per-instance load, affecting both latency and the GPU resources required to serve the workload.

As a large-scale agentic LLM service provider, we study these scheduling decisions through measurements of our production platform.
We characterize our agentic LLM workloads and analyze resource usage and latency across serving clusters.
This analysis identifies three requirements for efficient request scheduling.

\noindent\textbf{Requirement \#1: Preserving key--value cache reuse.}
Inference engines can retain previously computed attention keys and values in a key--value cache (KVC) for reuse by requests with matching prefixes~\cite{sglang2024,pensieve2025}.
This reuse is particularly valuable for agentic workloads, whose successive requests share long input prefixes~\cite{thunderagent2026}.
Our production trace shows a high input-to-output token ratio, with cached tokens accounting for approximately 80\% of input tokens (Fig.~\ref{fig:workflow-kvc-ttft}(a)).
Routing requests to instances that retain these prefixes avoids redundant prefill computation and saves GPU work~\cite{preble2025}.
Preserving this locality can also lower time-to-first-token (TTFT), the delay from request arrival to the first output token.

\noindent\textbf{Requirement \#2: Meeting workload-specific SLOs.}
Despite the high input-to-output token ratio, decode accounts for 78.4\% of measured LLM request duration in our production trace (Fig.~\ref{fig:workflow-request-duration}), making generation speed a key contributor to user-perceived latency.
Generation speed is measured by time-per-output-token (TPOT), the average interval between successive output tokens.
TPOT requirements vary across tasks: interactive coding may require fast generation, whereas background automation can tolerate slower responses~\cite{polyserve2025}.
Scheduling must therefore meet the TPOT service-level objective (SLO) configured for the target workload.

\noindent\textbf{Requirement \#3: Reducing the GPU footprint.}
Our production trace shows low per-instance request concurrency with substantial latency headroom.
Across five-minute samples, running concurrency ranges from 0.39 to 1.69 requests per instance, while mean TPOT increases only modestly and remains well below the SLO target (Fig.~\ref{fig:concurrency-tpot}).
This suggests room for \emph{request packing}: concentrating requests on fewer instances to increase decode batch sizes and improve throughput per GPU~\cite{sarathiserve2024}.
Scheduling should therefore use available TPOT headroom to serve the same demand with a smaller active GPU footprint while preserving the benefits of KVC reuse.

Existing schedulers generally adopt cache-aware or SLO-oriented approaches; however, jointly meeting the three requirements remains challenging (Table~\ref{tab:qualitative-comparison}).
First, cache-aware solutions combine prefix locality with instance load to avoid redundant prefill without creating hotspots~\cite{lmetric2026,dynamo2026}.
They do not explicitly exploit latency headroom to reduce the active GPU footprint.
Second, SLO-oriented systems use performance predictions to guide admission and scheduling~\cite{slosserve2025}, with some also packing requests onto fewer instances~\cite{polyserve2025,llmd2026}. 
However, for long-context agentic requests, missing cache locality can introduce significant prefill work to offset the gains from larger decode batches. 
The resulting interference can also delay ongoing decoding and jeopardize SLO attainment. 
The key challenge is therefore to minimize the active GPU footprint under SLO constraints by \emph{balancing the throughput gains from request consolidation against the prefill cost of lost cache reuse}.

In this paper, we present \systemname\systemnamefootnote, a gateway-level request scheduler that jointly addresses KVC reuse, TPOT SLO attainment, and GPU resource efficiency for agentic LLM serving.
We have deployed \systemname into our production clusters (see \S\ref{sec:production-evaluation}).
\systemname combines two core designs.

First, we develop a performance model for accurate, low-overhead latency prediction under prefill/decode interference.
Our analysis shows that relaxing the TPOT target permits larger decode batches, but the higher per-instance request rate also increases prefill occupancy, limiting the resulting throughput gain (Fig.~\ref{fig:capacity-validation}).
To capture these effects, we derive compact white-box models through offline profiling of prefill and decode under each deployment configuration.
During online scheduling, \systemname evaluates the calibrated models using request lengths, candidate-specific cached prefixes, and instance load to predict TTFT and TPOT.

Second, \systemname uses the performance model to guide request packing under SLO constraints based on two insights.
\emph{(1) KVC reuse should be prioritized over request packing.}
We observe that lost KVC reuse can cost more GPU time than larger decode batches save.
Accordingly, \systemname enforces a normalized recomputation budget that bounds each candidate's predicted excess prefill work relative to the minimum among capacity-eligible instances.
This budget controls how much cache locality can be traded for placement flexibility.
\emph{(2) Decode batch size is a key control variable for the throughput--latency tradeoff.}
For long-context agentic requests, each additional decoding request increases both the batch size and the aggregate KV state accessed per iteration.
Our measurements show that higher decode concurrency improves throughput per GPU but also raises TPOT, limiting how far requests can be packed (Fig.~\ref{fig:capacity-validation}).
\systemname therefore regulates decode concurrency through request placement.
Among cache-admissible instances predicted to meet the TPOT target, it favors those with more ongoing decoding requests, using cache locality and queue backlog to break ties.

We implement \systemname and evaluate it on a cluster of 64 NVIDIA H20 GPUs using production-derived agentic workloads.
Compared to LMetric and llm-d+, two state-of-the-art schedulers, \systemname uses 13.0--16.8\% and 21.1--24.6\% fewer active GPU-hours under 30-ms and 50-ms TPOT targets, respectively, while meeting the evaluated TPOT objectives.
Across both settings, its KVC hit rate is 79.5--80.2\%, close to LMetric's 81.0\% and higher than llm-d+'s 74.1--74.3\%.
We also compare two matched six-day windows (12 days total) on a production cluster comprising over 1000 GPUs.
Compared with a cache-aware load-balancing policy used in our production environment, \systemname achieves 34.7\% fewer serving instances and 36.8\% higher per-instance request throughput.

Our main contributions are summarized as follows:
\begin{itemize}
  \item We characterize production agentic workloads and identify three requirements for efficient scheduling: KVC reuse, TPOT SLO attainment, and GPU footprint reduction.
  \item We design and implement \systemname, a gateway-level scheduler that combines low-overhead performance prediction with SLO-aware request packing and bounded recomputation.
  \item We evaluate \systemname on a 64-GPU cluster, observing 13.0--24.6\% fewer active GPU-hours than LMetric and llm-d+ while meeting the evaluated windowed TPOT objectives. The 12-day production comparison further shows 34.7\% fewer serving instances and 36.8\% higher per-instance throughput.
\end{itemize}

\section{Background}
\label{sec:background}
\label{sec:related-work}

\paragraph{LLM inference and P/D colocation.}
Prefill builds the KVC from the prompt, while autoregressive decode reuses
and extends it to generate subsequent tokens~\cite{sarathiserve2024}.
In P/D-colocated serving, both stages share an instance's GPUs, so prefill
can lengthen mixed iterations or delay decode iterations, increasing token
latency for ongoing requests~\cite{sarathiserve2024}.

\paragraph{Cluster-level request scheduling.}
Cluster-level request scheduling routes each reques to a instance across serving clusters. 
Table~\ref{tab:qualitative-comparison} compares representative
scheduling mechanisms.
Load-based policies such as JSQ~\cite{winston1977jsq} and
vllm-dp~\cite{vllm2023} distribute requests according to instance load.
LMetric~\cite{lmetric2026} and Dynamo~\cite{dynamo2026} additionally account
for KVC reuse through multiplicative and weighted cache/load scores,
respectively.
SMetric~\cite{smetric2026} balances session-start requests and favors prefix
reuse for subsequent requests.
DualMap~\cite{dualmap2026} uses a TTFT target to balance cache affinity
against load, while llm-d~\cite{llmd2026} combines cache affinity with
TTFT/TPOT predictions to favor instances with small positive SLO headroom.
However, small TPOT headroom can reflect prefill interference rather than
larger decode batches (\S\ref{sec:routing-motivation}).

\paragraph{Instance-level request scheduling.}
Beyond selecting a destination instance, request scheduling can exercise
engine-level control over batch execution~\cite{orca2022,fastserve2026,vtc2024}, token allocation~\cite{jitserve2026}, or the
migration of ongoing requests. For example, 
Llumnix~\cite{llumnix2024} enables runtime rescheduling through live migration of requests
and their KVC across instances.
SLOs-Serve~\cite{slosserve2025} combines model-based admission and token allocation with request
rerouting.
PolyServe~\cite{polyserve2025} coordinates dynamic chunking and deadline-aware scheduling with
instance selection and autoscaling.
To limit integration and maintenance effort across inference engines in a
model-as-a-service (MaaS) platform, we focus on gateway-level request scheduling using
load and KVC information, without changing engine scheduling or migrating
ongoing requests.

\begin{table}[t]
\centering
\caption{Comparison with related work.}
\label{tab:qualitative-comparison}
\footnotesize
\setlength{\tabcolsep}{3pt}
\renewcommand{\arraystretch}{1.12}
\renewcommand{\tabularxcolumn}[1]{m{#1}}
\begin{tabularx}{\columnwidth}{@{}
  >{\raggedright\arraybackslash}X
  >{\centering\arraybackslash}m{0.105\columnwidth}
  >{\centering\arraybackslash}m{0.115\columnwidth}
  >{\centering\arraybackslash}m{0.135\columnwidth}
  >{\centering\arraybackslash}m{0.235\columnwidth}@{}}
\toprule
\textbf{Work}
  & \makecell[c]{\textbf{KVC}\\\textbf{reuse}}
  & \makecell[c]{\textbf{SLO-}\\\textbf{aware}}
  & \makecell[c]{\textbf{Request}\\\textbf{packing}}
  & \makecell[c]{\textbf{Performance}\\\textbf{model}} \\
\midrule
vllm-dp~\cite{vllm2023}
  & \no & \no & \no & -- \\
JSQ~\cite{winston1977jsq}
  & \no & \no & \no & -- \\
LMetric~\cite{lmetric2026}
  & \yes & \no & \no & -- \\
Dynamo~\cite{dynamo2026}
  & \yes & \no & \no & -- \\
DualMap~\cite{dualmap2026}
  & \yes & \yes & \no & White-box \\
SMetric~\cite{smetric2026}
  & \yes & \no & \no & -- \\
Llumnix~\cite{llumnix2024}
  & \no & \yes & \no & -- \\
SLOs-Serve~\cite{slosserve2025}
  & \no & \yes & \no & White-box \\
PolyServe~\cite{polyserve2025}
  & \no & \yes & \yes & Black-box \\
llm-d~\cite{llmd2026}
  & \yes & \yes & $*$ & Black-box \\
\textbf{\systemname}
  & \yes & \yes & \yes & \textbf{White-box} \\
\bottomrule
\end{tabularx}
\par\vspace{3pt}
\begin{minipage}{\columnwidth}
\footnotesize\raggedright
\yes{} and \no{} indicate whether a capability is explicitly supported.
White-box models predict performance with explicit modeling; black-box models use learned mappings or latency lookups.
A dash (--) indicates that no latency predictor is used.
$*$~Headroom-based packing in llm-d may not fully exploit the throughput benefits of larger decode batches (see \S\ref{sec:evaluation}).
\end{minipage}
\end{table}

\section{Characterization of Agentic Workloads}
\label{sec:workload-characterization}

We analyze a 14-day production trace of approximately 10.1 million
requests across 14 models, alongside a separate day-long trace of
instance concurrency and TPOT.
These observations motivate the three scheduling requirements
introduced in \S\ref{sec:introduction}.
Further analysis of these observations, together with controlled
experiments, yields two key insights that guide \systemname's design.

\begin{figure}[t]
  \centering
  \includegraphics[width=\columnwidth]{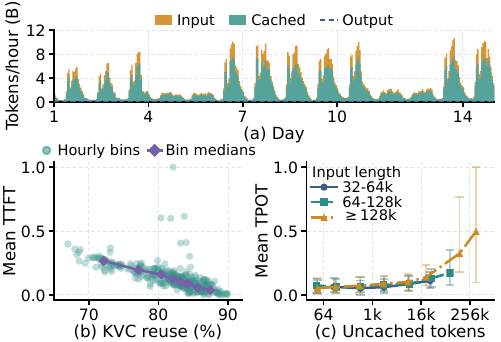}
  \caption{KVC reuse and latency. (a) Hourly tokens (cached input included). (b) Reuse versus TTFT. (c) Uncached input (log) versus background TPOT ($k=1024$). Latency axes are min--max normalized. }
  \label{fig:workflow-kvc-ttft}
\end{figure}

\phantomsection
\label{sec:workload-kvc}
\paragraph{Characterizing KVC reuse.}
Fig.~\ref{fig:workflow-kvc-ttft}(a) summarizes the hourly
input, cached-input, and output token volumes in the 14-day trace.
Input volume is approximately $192\times$ output volume, and cached
tokens account for 77.7\% of input tokens.
These requests are both input-heavy and rich in reusable context.
Successive calls in an agent session repeatedly include accumulated
execution history and tool results, allowing much of the input to reuse
previously computed KVC.
Recomputing these shared prefixes adds prefill work, which can delay
the first token and interfere with ongoing decoding.
For TTFT, we group hourly averages by KVC reuse.
The median is 66.2\% lower in the highest-reuse group than in the
lowest-reuse group (Fig.~\ref{fig:workflow-kvc-ttft}(b)).
We separately examine background TPOT, defined as the average
whole-request TPOT of other requests inferred to be decoding on the
same instance at a sampled arrival.
For inputs of at least 128k tokens, the group with the most uncached
tokens has about $2.2\times$ the median background TPOT of the group
with the fewest (Fig.~\ref{fig:workflow-kvc-ttft}(c)).
Together, these observations motivate preserving KVC reuse in request scheduling to limit prefill work and its impact on both TTFT and TPOT (\textbf{Requirement~\#1}).

\phantomsection
\label{sec:workload-slos}

\begin{figure}[t]
  \centering
  \includegraphics[width=\columnwidth]{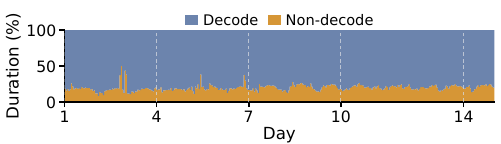}
  \caption{Request-duration composition over the 14-day trace.}
  \label{fig:workflow-request-duration}
\end{figure}

\paragraph{Characterizing decode demand.}
Although output tokens represent only a small fraction of token volume, decode accounts for 78.4\% of aggregate LLM request duration (Fig.~\ref{fig:workflow-request-duration}), approximately four times the remaining share.
Here, request-level decode duration includes waiting between tokens, rather than only GPU execution time.
The remainder includes prefill computation, queueing, network delays, and other gateway-side overhead.
This decode-dominated request duration motivates our focus on TPOT, as slower generation can delay subsequent calls in an agent's request chain.
We therefore adopt the TPOT SLO configured for the target workload as the latency constraint for online scheduling (\textbf{Requirement~\#2}; \S\ref{sec:resource-metrics}).

\begin{figure}[t]
  \centering
  \includegraphics[width=0.95\columnwidth]{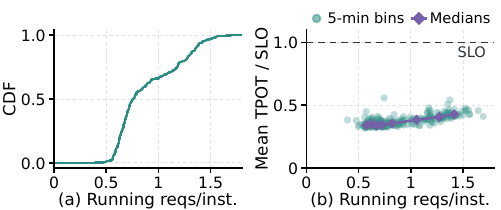}
  \caption{Mean running concurrency: (a) CDF and (b) TPOT relative to its SLO across five-minute samples. The dashed line marks the SLO.}
  \label{fig:concurrency-tpot}
\end{figure}

\phantomsection
\label{sec:routing-motivation}
\paragraph{Characterizing load distribution.}
To examine instance-level load and latency, we focus on a day-long production trace.
Fig.~\ref{fig:concurrency-tpot}(a) shows low average running concurrency, with a median of 0.76 requests per instance over the day.
Between the lowest and highest concurrency bins in Fig.~\ref{fig:concurrency-tpot}(b), median concurrency more than doubles, from 0.58 to 1.42 requests per instance, while the corresponding median of mean TPOT increases by only 25.6\%.
Low concurrency and this modest increase in TPOT suggest an opportunity to pack more requests onto each instance within a TPOT budget.
Such packing could improve resource utilization and reduce the GPU footprint (\textbf{Requirement~\#3}).

\paragraph{Capacity under prefill/decode colocation.}
\phantomsection
\label{sec:slo-capacity}
To explain the packing opportunity and its limits, we consider a homogeneous
steady load with arrival rate $\lambda$.  Requests share input, cached-prefix,
and output lengths $L_{\rm in}$, $L_{\rm cache}$, and $L_{\rm out}$.
The stage profiles in \S\ref{sec:profiling} provide prefill time $T_{\rm pre}$
and interference-free decode-iteration time $T_{\rm iter}(b)$ at mean decode
batch size $b$.  With mean active context length approximated by
$\bar{\ell}=L_{\rm in}+L_{\rm out}/2$, the engine's batch limit $B_{\max}$
and KV capacity $M_{\rm KV}$ (in tokens) bound feasible concurrency by
$b\leq b_{\max}=\min\{B_{\max},\lfloor M_{\rm KV}/\bar{\ell}\rfloor\}$.

Assuming prefill and decode occupy disjoint service time, approximately
$\lambda T_{\rm dec}$ new prefills interrupt a request's decode duration
$T_{\rm dec}$.  Thus,
$T_{\rm dec}=L_{\rm out}T_{\rm iter}(b)+\lambda T_{\rm dec}T_{\rm pre}$.
For sufficiently long outputs and $\lambda T_{\rm pre}<1$, this gives
\begin{equation}
  \operatorname{TPOT}_{\rm real}\approx
  \frac{T_{\rm iter}(b)}{1-\lambda T_{\rm pre}}.
  \label{eq:tpot-capacity}
\end{equation}
Applying Little's Law~\cite{little1961} to the decode phase gives
$b\approx\lambda L_{\rm out}\operatorname{TPOT}_{\rm real}$, and hence
\begin{equation}
  \lambda=\frac{b}{bT_{\rm pre}+L_{\rm out}T_{\rm iter}(b)}.
  \label{eq:capacity-arrival-rate}
\end{equation}
The pure-decode and wall-time output throughputs are therefore
\begin{equation}
  \mu_{\rm dec}=\frac{b}{T_{\rm iter}(b)},\qquad
  \mu_{\rm wall}=\lambda L_{\rm out}
  =(1-\lambda T_{\rm pre})\mu_{\rm dec}.
  \label{eq:capacity-throughputs}
\end{equation}
For a TPOT target $\tau$, we enumerate feasible $b$ and select the largest
batch satisfying the modeled $\operatorname{TPOT}_{\rm real}(b)\leq\tau$.
A larger batch can improve pure-decode throughput, but any accompanying
increase in arrival rate also raises prefill occupancy, limiting the
wall-time gain.  Prefix reuse reduces $T_{\rm pre}$ and leaves more service
time for decoding.  This steady-state approximation explains the tradeoff;
it is not a latency guarantee for heterogeneous, time-varying traffic.

\begin{figure}[t]
  \centering
  \includegraphics[width=\columnwidth]{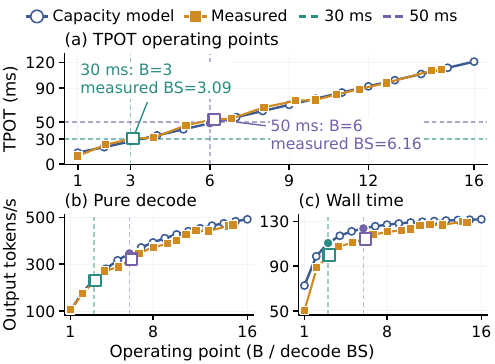}
  \caption{Decode batching under colocated prefill/decode. (a) TPOT,
  (b) pure-decode throughput, and (c) wall-time throughput.
  Lines compare the capacity model with measurements.}
  \label{fig:capacity-validation}
\end{figure}

A single-instance microbenchmark exhibits this trend
(Fig.~\ref{fig:capacity-validation}).  Between the measured points associated
with the 30- and 50-ms reference targets, decode batch size roughly doubles,
increasing pure-decode throughput by 40.4\% but wall-time throughput by only
14.4\%.  The smaller wall-time gain is consistent with increased prefill
occupancy.

\paragraph{Discussion and key insights.}
Together, the production observations and capacity analysis yield two
scheduling insights.

First, \emph{KVC reuse should be prioritized over request packing.}
Cache-dependent prefill work affects both first-token latency and ongoing
decoding (Fig.~\ref{fig:workflow-kvc-ttft}), reducing the throughput attainable
under a TPOT constraint.  Moving a request to an instance with a larger
decode batch need not save GPU time if additional recomputation erases the
decode gain.  Scheduling should therefore limit recomputation cost before
selecting among placements for packing.  This does not require maximizing
cache hits unconditionally; bounded locality loss can provide placement
flexibility.

Second, \emph{decode batch size is a key control variable for the
throughput--latency tradeoff.}
Larger batches can improve throughput per GPU, but increase the aggregate
KV state accessed per iteration and raise TPOT.  Queued and prefilling
requests do not contribute to the current decode batch, so total in-flight
counts alone cannot guide consolidation.  The gateway should favor
instances with more ongoing decoding requests under recomputation and
predicted TPOT constraints.

\section{System Overview}
\label{sec:system-overview}
\label{sec:system-design}

\begin{figure*}[t]
  \centering
  \includegraphics[width=\textwidth]{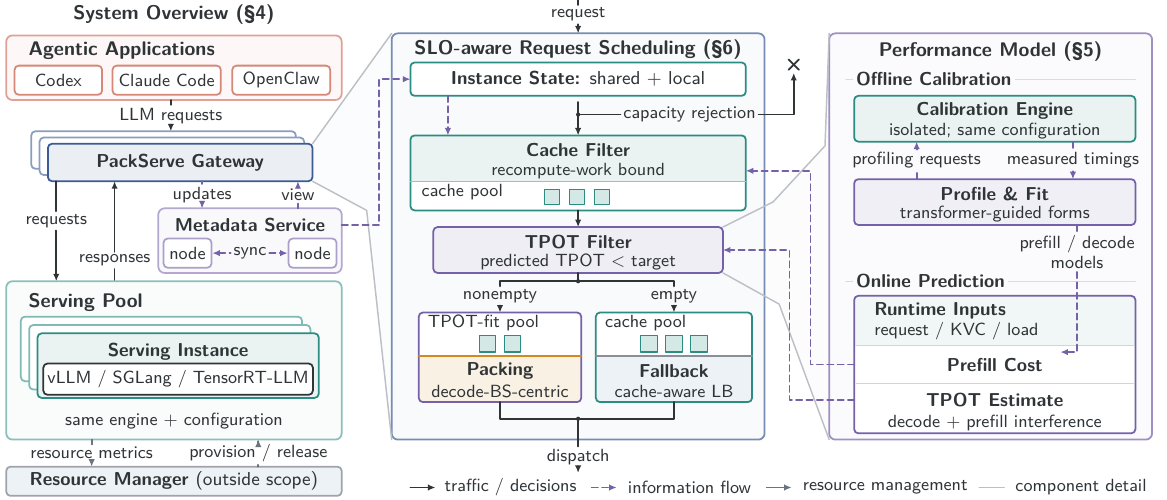}
  \caption{\systemname Overview: distributed gateways independently schedule
  requests to a shared serving pool using shared metadata.}
  \label{fig:system-overview}
\end{figure*}

Guided by the workload observations in \S\ref{sec:workload-characterization},
\systemname is a gateway-level scheduler designed to pack agentic LLM requests
under a configured TPOT objective while preserving KVC reuse.
As shown in Fig.~\ref{fig:system-overview}, distributed gateways route
requests from agent applications to a shared pool of serving instances,
with a Metadata service supplying instance state for scheduling. Each pool
contains homogeneous serving instances that use a common engine and
configuration~\cite{vllm2023,sglang2024,tensorrtllm2023}.
Engine-level scheduling and GPU provisioning remain outside the scheduler's
control.

\paragraph{Distributed gateways.}
Gateways carry critical inference traffic and scale horizontally to increase
request-handling capacity and maintain service availability as the cluster
grows. In a large production cluster, this can mean dozens of gateways.
Request scheduling is therefore distributed: each gateway independently selects
destinations for incoming requests from the shared serving pool.

\paragraph{Metadata service.}
These distributed decisions depend on timely per-instance load and cache
information. Because each gateway observes only its own traffic while an
instance serves requests from multiple gateways, stale state can cause load
or reusable prefixes to be misjudged. To maintain a shared view, gateways
track request concurrency, token usage, prefill load, and prefix-cache hits
through interactions along request and response paths rather than periodically
polling each engine. A logically centralized Metadata service aggregates this
per-instance state across gateways and makes it available for scheduling.
Multiple Metadata nodes synchronize for high availability, so gateways share
state while retaining independent scheduling decisions.

\paragraph{Scheduling policy.}
The scheduler within each gateway supports different scheduling policies, each using
request attributes and instance state to select a destination. Within this
framework, \systemname implements SLO-aware request packing to improve resource
efficiency while preserving KVC reuse. Its configuration-calibrated white-box
prefill and decode models (\S\ref{sec:prediction}) estimate recomputation cost
and TPOT: the prefill model estimates computation cost from new and cached
input tokens, while the TPOT estimate combines decode cost under estimated
load with prefill interference. After capacity filtering, a recomputation
budget bounds additional predicted prefill work, TPOT filtering retains
candidates predicted to meet the configured target, and a decode-batch-size-centric
selector favors candidates with more ongoing decoding requests
(\S\ref{sec:routing}).

\section{Performance Modeling}
\label{sec:prediction}

We derive prefill and decode models from Transformer computation and
calibrate them for each serving configuration (Fig.~\ref{fig:system-overview}).
These models estimate recomputation cost and TPOT for scheduling
(\S\ref{sec:routing}).

\subsection{Problem Definition}
\label{sec:modeling-goals}

\paragraph{Modeling goals.}
For an arriving request $r$ and candidate instance $i$, we estimate prefill
computation time and TPOT before scheduling, using only request attributes,
scheduler-observable instance state, and configuration-specific profiles.

\paragraph{TTFT decomposition.}
We decompose TTFT as
\begin{equation}
  \operatorname{TTFT}_{r,i}=
  T^{\rm pf}_{r,i}+T^{\rm queue}_{r,i}+T^{\rm overhead}_{r},
  \label{eq:ttft-decomposition}
\end{equation}
where $T^{\rm pf}$ is prefill computation time, $T^{\rm queue}$ is
queueing delay, and $T^{\rm overhead}$ covers scheduling and other
request-processing overheads.  Queueing delay can be estimated from request
state maintained by the scheduler, while scheduling overhead is negligible
relative to prefill in our setting.  We model cache-dependent $T^{\rm pf}$
to quantify candidate-specific recomputation cost and prefill interference.

\paragraph{TPOT decomposition.}
Under prefill/decode colocation, request-level TPOT comprises
interference-free decoding and prefill-induced delay:
\begin{equation}
  \operatorname{TPOT}_{r,i}=
  \overline{T}^{\rm dec}_{r,i}+\overline{\Delta}^{\rm PD}_{r,i},
  \label{eq:tpot-decomposition}
\end{equation}
where $\overline{T}^{\rm dec}$ is interference-free decode time and
$\overline{\Delta}^{\rm PD}$ is additional prefill-induced delay, both
averaged per output token.

\subsection{Transformer-Guided Stage Models}
\label{sec:stage-models}

\paragraph{Prefill.}
Transformer blocks~\cite{vaswani2017attention} combine causal self-attention
with token-wise projections and feed-forward or expert
layers~\cite{switchtransformers2022}.
For query, key, and value matrices $Q,K,V$, attention computes
\begin{equation}
  \operatorname{Attention}(Q,K,V)=
  \operatorname{softmax}\!\left(
    \frac{QK^{\mathsf T}}{\sqrt{d_h}}+M
  \right)V,
  \label{eq:causal-attention}
\end{equation}
where $d_h$ is the head dimension and $M$ masks future positions.

With $C$ cached prefix tokens and $N$ new tokens, causal masking permits
prefix K/V reuse, leaving token-wise computation only for the new tokens.
Each new query attends to all cached keys and new keys up to and including
its own position, yielding $NC+N(N+1)/2$ query-key pairs per head.

For a fixed configuration $\theta$ (model, hardware, and engine settings),
these costs motivate
\begin{equation}
  T_{\rm pf}(N,C;\theta) \approx
  \alpha_0+\alpha_1N+\alpha_2NC+\alpha_3N^2,
  \label{eq:prefill-structural}
\end{equation}
where the linear term primarily captures token-wise computation, while $NC$
and $N^2$ capture attention interactions.  Distinguishing new and cached
tokens captures how a longer reusable prefix reduces $N$ and recomputation
at fixed input length.

\paragraph{Decode.}
Each standard autoregressive decode iteration generates one token for each
of $B$ active requests with total context length
$S=\sum_{j=1}^{B}\ell_j$.  Token-wise operations process $B$ new tokens,
while attention accesses KVC proportional to $S$, motivating the local
approximation
\begin{equation}
  T_{\rm iter}(B,S;\theta) \approx
  \beta_0+\beta_1B+\beta_2S,
  \label{eq:decode-model}
\end{equation}
where coefficients are calibrated per configuration.  The model excludes
prefill interference.

\subsection{Configuration-Calibrated Profiling}
\label{sec:profiling}

We profile each serving configuration on an otherwise idle instance, within
its context-length, KVC-capacity, and concurrency limits.

\paragraph{Prefill profile.}
We add empirical $C$ and $C^2$ terms to capture cache-dependent runtime effects
beyond the structural model.  Sampling covers the cold-cache axis ($C=0$), chunk-aligned
cache-hit points ($C=kL_{\mathrm{chunk}}$, $0<N\le L_{\mathrm{chunk}}$), and
the maximum-input boundary ($C+N=L_{\max}$), where $L_{\mathrm{chunk}}$ is the
prefill chunk size, $k$ is a positive integer, and $L_{\max}$ is the maximum
feasible input length.  For cache-hit samples, we prewarm the prefix and
verify the cached-token count reported by the engine.  Each profiling
request generates only one output token.

\paragraph{Decode profile.}
We vary request concurrency and output length to sample $(B,S)$.  When
available, we collect actual $B$, $S$, and execution time from engine-native
forward-pass metrics (FPM), retaining only pure-decode batches.  Otherwise,
we estimate iteration intervals from streamed output timestamps.  After
outlier filtering, we fit separate linear models using CUDA graph capture
batch sizes as segment boundaries.

\paragraph{Online calibration.}
\phantomsection
\label{sec:online-calibration}
Profiled decode latency can underestimate runtime cost at large batch sizes,
potentially due to context-length skew~\cite{vidur2024,llmservingsim22025}.
To reduce this bias, \systemname maintains a multiplicative exponential
moving average (EMA) factor for each decode-BS segment.  Feedback uses
engine-native durations of consecutive pure-decode forwards and the base
model's predictions at the same measured batch size and total context
length, excluding prefill interference from the correction.
Each factor starts at 1 and tracks the measured-to-predicted ratio with
update weight 0.02; ratios are clipped to $[0.5,2]$ and factors to $[1,2]$.
At scheduling time, the factor for the projected batch size scales the
decode term, while prefill interference is modeled separately.
We evaluate the correction in \S\ref{sec:predictor-comparison}.

\section{SLO-Aware Request Scheduling}
\label{sec:routing}

Using the performance models in \S\ref{sec:prediction}, \systemname
balances the throughput gains from request packing against the prefill
cost of lost KVC reuse.
Algorithm~\ref{alg:slo-aware-routing} implements three stages motivated 
by the insights in \S\ref{sec:workload-characterization}: 
KVC-first filtering, SLO-aware admission, and decode-batch-size-centric packing.

\begin{algorithm}[t]
\caption{KVC-first, SLO-aware request packing.}
\label{alg:slo-aware-routing}
\small
\textbf{Input:} request $r$, instance pool and state $\mathcal{I}$,
recomputation budget $\eta\ge0$, and TPOT target $\tau^{\rm TPOT}$.\par
\textbf{Output:} selected instance $i^*$, or \textsc{Reject}.\par
\vspace{2pt}
\begin{tabularx}{\columnwidth}{@{}r@{\hspace{0.5em}}>{\raggedright\arraybackslash}X@{}}
\multicolumn{2}{l}{\textit{\# Capacity guard}} \\
\algline{line:capacity-candidates} & $\mathcal{A}\gets\textsc{CapacityCandidates}(r,\mathcal{I})$ \\
\algline{line:capacity-reject} & \textbf{if} $\mathcal{A}=\varnothing$ \textbf{then return} \textsc{Reject} \\
\multicolumn{2}{l}{\textit{\# KVC-first recomputation filtering}} \\
\algline{line:recompute-loop} & \textbf{for each} $i\in\mathcal{A}$ \textbf{do} \\
\algline{line:recompute-cost} & \hspace*{1em}$W_{r,i}\gets\textsc{RecomputeCost}(r,i)$ \\
\algline{line:recompute-min} & $W_{\min}\gets\min_{i\in\mathcal{A}} W_{r,i}$ \\
\algline{line:cache-benefit} & $Z_r\gets\textsc{CacheBenefit}(r)$ \\
\algline{line:recompute-filter} & $\mathcal{C}\gets\{i\in\mathcal{A}\mid
     \textsc{Normalize}(W_{r,i}-W_{\min},Z_r)\le\eta\}$ \\
\multicolumn{2}{l}{\textit{\# SLO-aware admission}} \\
\algline{line:tpot-loop} & \textbf{for each} $i\in\mathcal{C}$ \textbf{do} \\
\algline{line:tpot-predict} & \hspace*{1em}$\widehat{t}_i\gets\textsc{PredictTPOT}(r,i)$ \\
\algline{line:tpot-filter} & $\mathcal{T}\gets\{i\in\mathcal{C}\mid
      \widehat{t}_i<\tau^{\rm TPOT}\}$ \\
\multicolumn{2}{l}{\textit{\# Decode-Batch-Size-Centric packing or bounded fallback}} \\
\algline{line:packing-if} & \textbf{if} $\mathcal{T}\ne\varnothing$ \textbf{then} \\
\algline{line:packing-loop} & \hspace*{1em}\textbf{for each} $i\in\mathcal{T}$ \textbf{do} \\
\algline{line:packing-key} & \hspace*{2em}$K_i\gets\textsc{DecodeBatchSizeCentricKey}(r,i)$ \\
\algline{line:packing-select} & \hspace*{1em}$i^*\gets\arg\min_{i\in\mathcal{T}}^{\rm lex} K_i$ \\
\algline{line:fallback-else} & \textbf{else} \\
\algline{line:fallback-select} & \hspace*{1em}$i^*\gets\textsc{CacheAwareLB}(r,\mathcal{C})$ \\
\algline{line:routing-return} & \textbf{return} $i^*$ \\
\end{tabularx}
\end{algorithm}

\subsection{KVC-First Recomputation Control}
\label{sec:capacity-cache-admission}

After capacity filtering (\mbox{L\ref{line:capacity-candidates}--\ref{line:capacity-reject}}),
\systemname prioritizes KVC reuse, as motivated by \textit{Insight \#1} (\S\ref{sec:workload-kvc}).
The same cache-hit ratio can entail different recomputation costs depending
on input length and serving configuration.
Rather than applying a fixed cache-hit-ratio threshold, \systemname filters
candidates using predicted recomputation cost, accounting for the request's
input length, candidate-specific prefix reuse, and configuration-calibrated
prefill performance
(\mbox{L\ref{line:recompute-loop}--\ref{line:recompute-filter}}).

Specifically, let $\mathcal{A}$ denote the capacity-eligible instances.
For request $r$, $W_{r,i}$ is candidate $i$'s nonnegative predicted prefill-time
penalty relative to a fully cached input, and $W_{\min}$ is its minimum over
$\mathcal{A}$. Let $Z_r$ denote the predicted prefill-time difference between
fully uncached and fully cached inputs.
The KVC-admissible set $\mathcal{C}$ retains candidates satisfying
\begin{equation}
  \rho_{r,i}=\frac{W_{r,i}-W_{\min}}{Z_r}\le\eta,
  \qquad Z_r>0,
  \label{eq:normalized-recompute}
\end{equation}
where $\eta$ bounds additional recomputation relative to the best candidate,
expressed as a fraction of the request's full cache benefit.
For $Z_r>0$, $\eta=0$ retains only minimum-cost candidates, while larger values
permit more flexibility for subsequent packing (\S\ref{sec:eta-sensitivity}).
If $Z_r=0$, all capacity-eligible candidates remain.

\subsection{Decode-Batch-Size-Centric Packing}
\label{sec:tpot-admission}
\label{sec:consolidation-overload}

Passing the recomputation filter does not ensure that a placement
meets the TPOT target.
An arriving request adds both decode load and cache-dependent
prefill work to its destination.
For each KVC-admissible instance $i\in\mathcal{C}$, \systemname
therefore evaluates a placement-dependent TPOT estimate using
the performance models in \S\ref{sec:prediction}
(\mbox{L\ref{line:tpot-loop}--\ref{line:tpot-filter}}):
\begin{equation}
  \widehat{\operatorname{TPOT}}_{r,i}=
  \widehat{T}_{\rm iter}(B_i^+,S_{r,i}^+)
  +\mathbf{1}[A_i>0]\,
   \frac{\widehat{T}_{\rm pf}(n_{r,i},c_{r,i})}
        {\max(1,\widetilde{L}_{\rm out})}.
  \label{eq:routing-tpot}
\end{equation}
Here, $A_i$ counts scheduler-tracked active requests, including queued
requests, and $B_i^+=A_i+1$ estimates potential decode concurrency
after admission rather than the currently executing batch size.
$S_{r,i}^+$ estimates the corresponding total context length.
The second term accounts for potential interference from the
arriving request's prefill, using its uncached and cached input
lengths $n_{r,i}$ and $c_{r,i}$.
We amortize this cost over the recent median output length
$\widetilde{L}_{\rm out}$ and omit it when the instance has no
active requests.
The decode term includes the BS-specific online correction described in
\S\ref{sec:online-calibration}.
Candidates whose estimated TPOT is below the pool's configured
target $\tau^{\rm TPOT}$ form the admissible set $\mathcal{T}$.

Within $\mathcal{T}$, \systemname prioritizes instances with more
scheduler-tracked decoding requests
(\mbox{L\ref{line:packing-if}--\ref{line:packing-select}}).
This preference encourages larger decode batches to improve
throughput per GPU, as motivated by \textit{Insight \#2}
(\S\ref{sec:routing-motivation}).
Selecting the smallest TPOT headroom would not necessarily
identify such instances, because small headroom can also result
from long contexts or prefill interference.
We therefore use predicted TPOT to constrain placement and
decode concurrency to guide packing.
When decoding concurrency is equal, \systemname prefers more
cached prefix tokens.
This tie-break preserves additional KVC reuse because the
recomputation filter bounds cache loss without making all
candidates equally cache-efficient.
Remaining ties favor fewer queued requests to limit queueing delay.

If $\mathcal{T}$ is empty, no KVC-admissible instance is predicted
to meet the TPOT target.
\systemname then switches from decode-batch-size-centric packing to cache-aware
load balancing within $\mathcal{C}$
(\mbox{L\ref{line:fallback-else}--\ref{line:fallback-select}}).
This fallback selects according to load and cache locality
rather than prioritizing further decode consolidation.
It retains the capacity and recomputation constraints but
relaxes the TPOT SLO test.

\section{Implementation}
\label{sec:implementation}

We build an experimental prototype of \systemname based on the scheduling
architecture of our existing production inference gateway to evaluate the
proposed scheduling mechanisms under controlled conditions.  Its lightweight
scheduling and performance-model modules comprise approximately 1.4K LOC of
Python, supported by 2.6K LOC of shared scheduler infrastructure.  Following
the Metadata service design, the prototype maintains a local prefix tree
to estimate KVC reuse and per-instance backend records for request counts
and load estimates.  These structures are updated along request and
response paths to support the performance models and scheduling policy.
To keep the implementation simple and efficient, we use
LMetric~\cite{lmetric2026} for the cache-aware load-balancing fallback
(\S\ref{sec:consolidation-overload}).
A separate profiler, implemented in approximately 2.8K LOC of Go, collects
prefill and decode measurements to calibrate the models.  In production,
we integrate our \systemname scheduling policy into the existing gateway
infrastructure.  Both implementations follow the same design principles
(\S\ref{sec:system-design}).

\section{Evaluation}
\label{sec:evaluation}

\ifearlyeoverview
\begin{figure*}[t]
  \centering
  \includegraphics[width=0.96\textwidth]{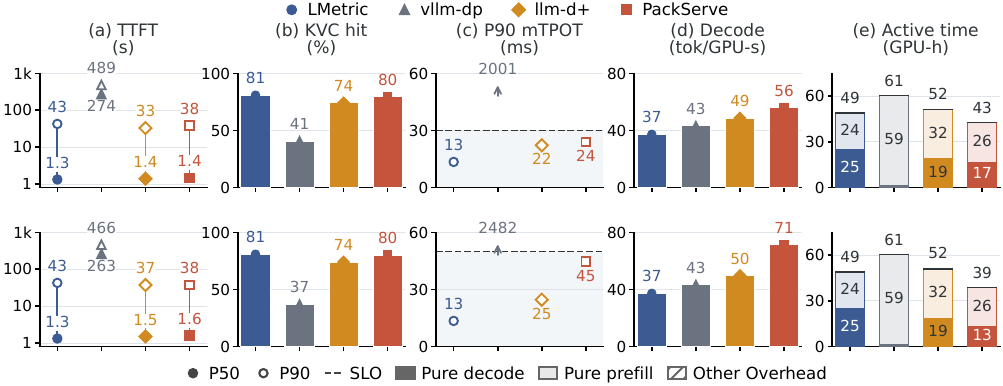}
  \caption{End-to-end performance under 30/50-ms TPOT SLOs (top/bottom).
  All policies except vllm-dp exceed 99.9\% request success.
  Most values are rounded to integers for clarity.}
  \label{fig:e2e-overview}
\end{figure*}
\fi

We evaluate \systemname on a 64-GPU testbed using a production-derived
agentic workload.  A separate 12-day production study examines a scheduling
policy sharing \systemname's design principles.  Our evaluation addresses
three questions:

\begin{itemize}
\item \textbf{End-to-end effectiveness:} Can \systemname reduce active
GPU-hours relative to existing scheduling policies while retaining KVC reuse
and meeting windowed TPOT objectives? (\S\ref{sec:end-to-end})
\item \textbf{Design analysis:} How do \systemname's scheduling components
affect resource use, latency, and KVC reuse, and how does online calibration
change decode-model error? (\S\ref{sec:ablation})
\item \textbf{Production observations:} What changes in serving-instance
count, per-instance throughput, latency, and KVC reuse accompany SLO-aware
scheduling in production? (\S\ref{sec:production-evaluation})
\end{itemize}

\subsection{Experimental Setup}
\label{sec:resource-metrics}
\label{sec:resource-boundary}

\paragraph{Testbed.}
We use a Kubernetes cluster with 64 NVIDIA H20 GPUs (96~GB each) and
intra-node NVLink connectivity.  We deploy 16 homogeneous
SGLang~\cite{sglang2024} serving instances, each using 4 GPUs within a
single node with tensor parallelism of 4.  The scheduler runs in a dedicated
CPU-only pod.

\paragraph{Model and serving configuration.}
All instances serve 230B MiniMax-M2.5~\cite{minimax2026m2series}
with colocated prefill and decode.  We use
8192-token prefill chunks, a maximum batch size of 64, and FP8 KVC with
approximately one million tokens of GPU-resident capacity per instance.
HiCache adds a host-memory cache tier configured to $3\times$ the
GPU-cache capacity.

\paragraph{Production-derived workload.}
We construct a one-hour workload from anonymized production traffic served
by a roughly 700B-parameter open-weight model.  We sample complete sessions 
to obtain a steady load while preserving within-session request intervals.
Table~\ref{tab:workload-summary} summarizes the resulting workload.
Each run uses the first 10 minutes to warm caches and scheduler history,
and evaluates requests arriving during the remaining 50 minutes.

\begin{table}[t]
  \centering
  \caption{Replay workload characteristics, including warmup.}
  \label{tab:workload-summary}
  \small
  \begin{tabular*}{\columnwidth}{@{\extracolsep{\fill}}lrlr@{}}
    \toprule
    \textbf{Workload} & \textbf{Value} & \textbf{Mean length} & \textbf{Tokens} \\
    \midrule
    Requests & 7,258 & Input & 62.5K \\
    Sessions & 1,704 & Output & 556 \\
    Mean qps & 2.02~req/s & Reusable prefix & 52.1K \\
    \bottomrule
  \end{tabular*}
  \par\smallskip
  \begin{minipage}{\columnwidth}
    \footnotesize
    Reusable prefix: global, unbounded cache assumed.
  \end{minipage}
\end{table}

\paragraph{Baselines.}
We compare \systemname with three baselines from the design space in
Table~\ref{tab:qualitative-comparison}, adapting and reproducing their
scheduling policies in our experimental prototype:
\begin{itemize}
\item \textbf{vllm-dp}~\cite{vllm2023} adapts vLLM v0.28.0's default
internal DP load balancer. It selects the instance with the lowest load
score, combining active request count with a prefill-backlog term weighted
by KVC occupancy above 50\%. It does not consider request-specific KVC
reuse or latency SLOs.
\item \textbf{LMetric}~\cite{lmetric2026} balances KVC reuse and load by
selecting the instance that minimizes the product of pending uncached
prefill tokens and active request count, both including the arriving
request. We track pending prefill work at the gateway until the first
token returns.
\item \textbf{llm-d+} adapts llm-d v0.9.0~\cite{llmd2026}, which combines
KVC affinity with latency-based request packing. It applies cache-affinity
checks and prioritizes instances predicted to satisfy both TTFT and TPOT
SLOs, then uses weighted random selection to favor smaller positive
latency headroom. We replace its XGBoost predictors with our profile-based
models and online calibration for better TPOT prediction accuracy
(\S\ref{sec:predictor-comparison}), denoting this adaptation as llm-d+.
\end{itemize}

\paragraph{Metrics.}
We evaluate request success, latency, KVC reuse, and GPU resource use
using the following metrics.
\begin{itemize}
\item \textit{Request success rate.}
We report the fraction of all evaluation requests that complete successfully.
A request is considered failed if no instance can admit it due to capacity
limits or it receives no response within 10 minutes.

\item \textit{TTFT.}
We report client-observed TTFT P50/P90.
Rather than using a fixed workload-wide
threshold~\cite{distserve2024,mooncake2025}, we set each request's TTFT SLO
to the pure-prefill computation time for its full input, with no KVC reuse
or interference from other requests.  We estimate this time through offline
profiling (\S\ref{sec:profiling}) and impose a minimum budget of 2\,s to
leave slack for short inputs.  We prioritize KVC reuse on the prefill side,
using TTFT as a reference metric to assess increases in recomputation
and queueing overhead. Our llm-d+ adaptation
uses the same profile-based budget and 2-s minimum for scheduling.

\item \textit{TPOT.}
We evaluate TPOT SLOs of 30 and 50~ms in separate runs, motivated by
the distinct decode-throughput operating points in our trace-derived
homogeneous microbenchmark (Fig.~\ref{fig:capacity-validation}).
Prefill interference under P/D colocation complicates request-tail control,
so we focus on P50 TPOT to characterize typical performance.
To capture temporal variation, we track \emph{minute-median TPOT (mTPOT)}:
the P50 TPOT of successfully completed requests issued within each
non-overlapping one-minute window.
We report the mTPOT time series and its P90 across valid windows.

\item \textit{KVC hit.}
We report the token-weighted cache-hit rate, computed as the total cached
input tokens divided by the total input tokens across successfully
completed evaluation requests.

\item \textit{GPU resource use.}
We report active GPU-hours, summing each instance's time with at least
one active request (including failed ones) weighted by its GPU count.
We also report successful decode throughput as output tokens after the
first token from successful requests per GPU-second of pure-decode service.

\end{itemize}

\ifearlyeoverview\else
\begin{figure*}[t]
  \centering
  \includegraphics[width=0.96\textwidth]{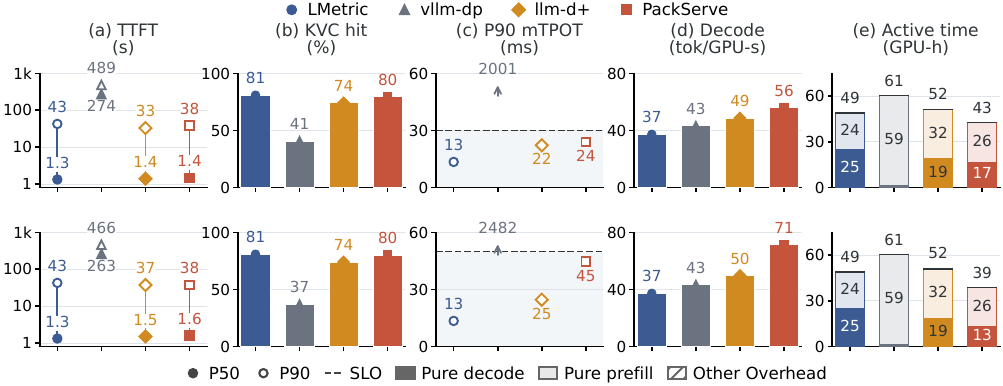}
  \caption{End-to-end performance under 30/50-ms TPOT SLOs (top/bottom).
  All policies except vllm-dp exceed 99.9\% request success.
  Most values are rounded to integers for clarity.}
  \label{fig:e2e-overview}
\end{figure*}
\fi

\iflateeplots\else
\begin{figure}[t]
  \centering
  \includegraphics[width=\columnwidth]{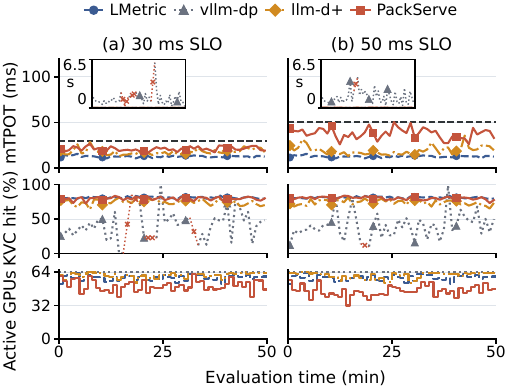}
  \caption{\systemname preserves cache locality with fewer active GPUs. Insets show vllm-dp's full mTPOT range; dashed lines mark SLOs. Red crosses mark all-failed vllm-dp arrival minutes.}
  \label{fig:e2e-timeseries-paired}
\end{figure}

\begin{figure}[t]
  \centering
  \includegraphics[width=\columnwidth]{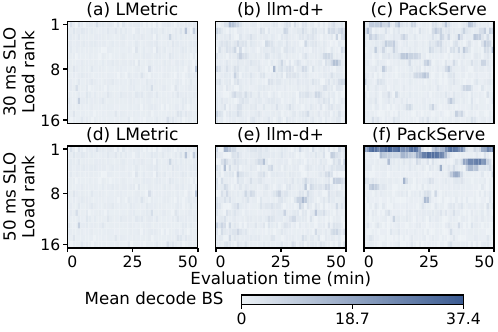}
  \caption{Decode batch sizes across load-ranked instances.}
  \label{fig:decode-bs-heatmap}
\end{figure}
\fi

\subsection{End-to-End Performance}
\label{sec:end-to-end}

Fig.~\ref{fig:e2e-overview} summarizes end-to-end performance.
Figs.~\ref{fig:e2e-timeseries-paired} and~\ref{fig:decode-bs-heatmap}
show temporal behavior and decode-batch distributions. We use
$\eta=0.5$ by default.
All policies except vllm-dp complete over 99.9\% of evaluation requests
in both settings, while vllm-dp completes only ~24\% of requests. 
Its low success rate and high TTFT and mTPOT
highlight the limits of load balancing without request-specific KVC awareness.
We therefore focus on LMetric and llm-d+ in the analysis
but retain vllm-dp in the figures unless otherwise noted.

\paragraph{Latency and SLO attainment.}
\systemname controls TPOT through explicit decode/interference prediction
and SLO-based candidate filtering. Its P90 mTPOT remains below both SLOs,
at 24.0 and 44.7\,ms (Fig.~\ref{fig:e2e-overview}(c)). Over time, mTPOT
follows the configured latency scale and exceeds it in only one 50-ms
minute by 0.3\,ms (Fig.~\ref{fig:e2e-timeseries-paired}, top row).
In contrast, LMetric stays at 10.5--15.6\,ms, consistent with its
cache-aware load balancing for latency minimization. llm-d+ also meets
every minute-level target, but its P90 mTPOT of 22.2 and 24.6\,ms leaves
more margin unused at 50\,ms. Despite sharing our predictors, its cache
affinity and joint TTFT/TPOT scoring do not regulate TPOT toward the
configured target. These results support explicit decode/interference
modeling and TPOT filtering for minute-level latency control.

\iflateeplots
\suppressfloats[t]
\begin{figure}[t]
  \centering
  \includegraphics[width=\columnwidth]{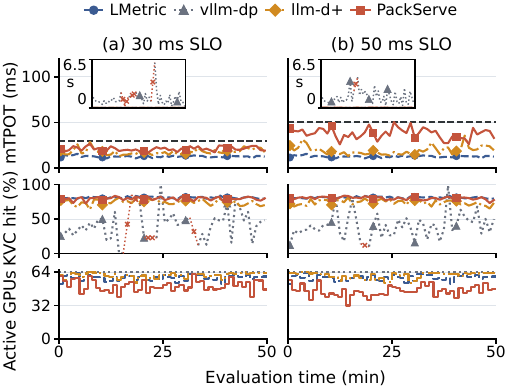}
  \caption{\systemname preserves cache locality with fewer active GPUs. Insets show vllm-dp's full mTPOT range; dashed lines mark SLOs. Red crosses mark all-failed vllm-dp arrival minutes.}
  \label{fig:e2e-timeseries-paired}
\end{figure}
\begin{figure}[t]
  \centering
  \includegraphics[width=\columnwidth]{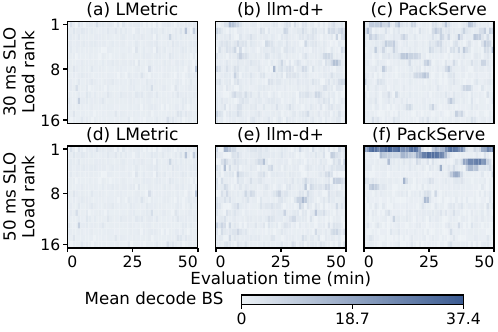}
  \caption{Decode batch sizes across load-ranked instances.}
  \label{fig:decode-bs-heatmap}
\end{figure}
\fi

Packing can increase recomputation and queueing, creating a TTFT tradeoff.
Nevertheless, \systemname's KVC-aware admission keeps median TTFT close
to LMetric's, at 1.4--1.6 versus 1.3\,s, while achieving a lower P90 of
37.6--38.4 versus 42.8\,s (Fig.~\ref{fig:e2e-overview}(a)). llm-d+ has
lower P90 TTFT at 30\,ms (32.9 versus 38.4\,s), with little difference
at 50\,ms. This is consistent with its joint objective and default
0.8/0.2 weights for normalized TTFT/TPOT headroom. \systemname instead
prioritizes KVC reuse under its TPOT SLO, using TTFT as a reference metric
to assess increases in recomputation and queueing overhead.

\paragraph{KVC reuse during consolidation.}
\systemname uses a recomputation budget to limit the locality cost of
concentrating requests. Its KVC hit rate remains within 1.5 percentage
points (pp) of LMetric's (Fig.~\ref{fig:e2e-overview}(b)), and their curves stay close
over time (Fig.~\ref{fig:e2e-timeseries-paired}, middle row), supporting
cache-first admission during consolidation. \systemname outperforms
llm-d+ by 5.3--6.1\,pp overall and remains ahead in 96--100\% of replay
minutes. llm-d+ uses fixed prefix-affinity thresholds of 0.99 (strict)
and 0.80 (loose), relaxed according to TTFT. These may require
workload-specific tuning because the same hit ratio can imply different
recomputation costs. In contrast, \systemname's profiled prefill model
estimates request-specific cost for the serving configuration, accounting
for input and cached-prefix lengths without a fixed hit-rate threshold.

\paragraph{GPU resource efficiency.}
\systemname reduces active GPU-hours by 13.0--21.1\% relative to LMetric
and 16.8--24.6\% relative to llm-d+ (Fig.~\ref{fig:e2e-overview}(e)).
It also uses fewer active GPUs than either baseline in at least 96\% of
replay minutes, showing that the savings persist over time
(Fig.~\ref{fig:e2e-timeseries-paired}, bottom row). 

Within the cache- and TPOT-admissible set, \systemname prioritizes
decode concurrency rather than smaller SLO headroom. Its decode throughput
is 50.1--90.8\% higher than LMetric's, consistent with the batching trend
in \S\ref{sec:routing-motivation} (Fig.~\ref{fig:e2e-overview}(d)).
Both its pure-decode and pure-prefill GPU-hours are lower than llm-d+'s,
with 18.4--21.1\% less prefill time (Fig.~\ref{fig:e2e-overview}(e)).
Whereas llm-d+'s
small-headroom objective can favor prefill work over larger decode batches,
\systemname improves batching within a recomputation budget.

Fig.~\ref{fig:decode-bs-heatmap} shows the resulting distribution:
LMetric spreads small decode batches relatively evenly, llm-d+ remains
broadly distributed, and \systemname concentrates larger batches on fewer
instances. This is clearest at 50\,ms, where high-load instances sustain
large batches over time, supporting decode-batch-size-centric packing as 
a way to reduce active capacity-time rather than provisioned GPU allocation.

\subsection{Ablation Study}
\label{sec:ablation}

We examine admission and fallback, the recomputation budget, TPOT prediction
and calibration, and instance selection.

\subsubsection{Layer-by-Layer Admission Ablation}
\label{sec:layer-ablation}

\begin{figure}[t]
  \centering
  \includegraphics[width=\columnwidth]{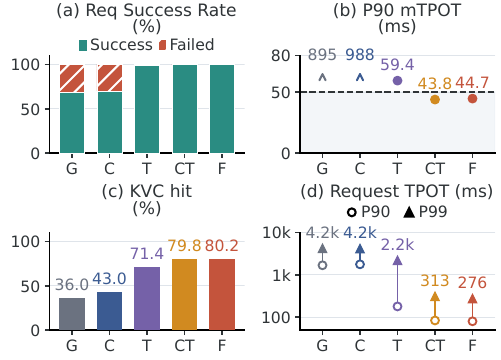}
  \caption{Admission and fallback ablation (50\,ms SLO). Arrows in (b) mark off-scale values; (d) uses a log scale.}
  \label{fig:layer-ablation}
  \par\medskip
  \begin{minipage}{\linewidth}
\makeatletter
\def\@captype{table}
\makeatother
\caption{Configuration key for Fig.~\ref*{fig:layer-ablation}.}
\label{tab:layer-ablation-configurations}
\centering
\begingroup
\fontsize{8}{9.5}\selectfont
\setlength{\tabcolsep}{3pt}
\renewcommand{\arraystretch}{1.0}
\begin{tabular*}{\linewidth}{@{\extracolsep{\fill}}llccc@{}}
\toprule
\textbf{ID} & \textbf{Configuration} & \textbf{Cache} & \textbf{TPOT} & \textbf{Fallback} \\
\midrule
G & Capacity guards only & $-$ & $-$ & -- \\
C & Cache only & $+$ & $-$ & -- \\
T & TPOT only & $-$ & $+$ & DCQ \\
CT & Cache + TPOT & $+$ & $+$ & DCQ \\
F & Full & $+$ & $+$ & Cache-aware LB \\
\bottomrule
\end{tabular*}
\par\smallskip
\raggedright
$+/-$: admission filters on/off. All retain capacity guards, prefix caching and normal DCQ selection. Fallback pools: capacity for T; cache for CT/F.
\par\endgroup
\end{minipage}

\end{figure}

We vary cache/TPOT admission and fallback under the 50\,ms setup of
\S\ref{sec:end-to-end}, using the configurations in
Table~\ref{tab:layer-ablation-configurations}.

\paragraph{TPOT admission and overload control.}
TPOT admission makes predicted latency an explicit scheduling constraint
(\S\ref{sec:tpot-admission}).
Capacity guards alone (G) and cache-only admission (C) incur failure
rates of 29.9--31.9\%.
TPOT-only admission (T) lowers the failure rate to 0.8\%
(Fig.~\ref{fig:layer-ablation}(a)), 
and it also reduces P90 mTPOT from 895.0--987.7\,ms to 59.4\,ms
(Fig.~\ref{fig:layer-ablation}(b)).
This remains above the 50\,ms SLO.
These results support model-based TPOT admission as the primary
overload control, beyond capacity and cache bounds alone.

\paragraph{KVC reuse and prefill interference.}
\systemname's cache-first design complements TPOT admission by bounding
additional prefill work.
Compared with T, combined cache-and-TPOT admission (CT) improves the
KVC hit rate by 8.4\,pp (Fig.~\ref{fig:layer-ablation}(c)) and reduces
P90 mTPOT to 43.8\,ms, below the SLO
(Fig.~\ref{fig:layer-ablation}(b)).
Its request-level P99 TPOT is also 85.9\% lower
(Fig.~\ref{fig:layer-ablation}(d)).
The joint improvement in locality and latency is consistent with
avoiding cache-destructive placements that introduce additional prefill
interference (\S\ref{sec:capacity-cache-admission}).

\paragraph{Fallback and tail latency.}
When no candidate passes TPOT admission, \systemname uses a
cache-aware load-balancing fallback over the cache-admissible pool,
balancing queued and incoming prefill work against active concurrency (\S\ref{sec:consolidation-overload}).
Full (F) retains a P90 mTPOT of 44.7\,ms, below the SLO
(Fig.~\ref{fig:layer-ablation}(b)).
Compared with CT, its request-level P99 TPOT is 11.6\% lower
(Fig.~\ref{fig:layer-ablation}(d)), while its request-level TPOT SLO
attainment is 1.7\,pp lower.
The lower observed extreme tail is consistent with load-aware fallback,
but does not establish uniform latency improvement.

\subsubsection{Impact of the Recomputation Budget}
\label{sec:eta-sensitivity}

We sweep $\eta$ from 0 to 1 in steps of 0.1 using the 50\,ms replay setup
of \S\ref{sec:end-to-end}, with all other strategy parameters fixed.

\begin{figure}[t]
  \centering
  \includegraphics[width=\columnwidth]{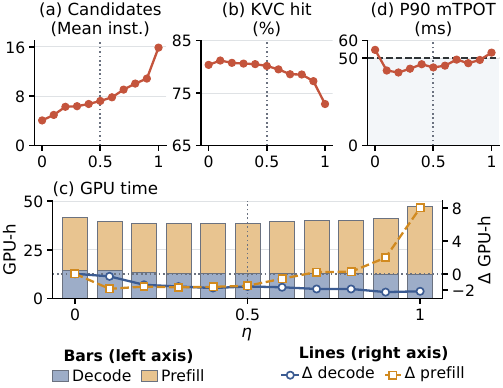}
  \caption{Impact of the recomputation budget. In (c), bars show GPU time (left axis); lines show changes from $\eta=0$ (right axis).}
  \label{fig:eta-sensitivity}
\end{figure}

\paragraph{Placement flexibility and cache locality.}
The normalized budget controls how much predicted recomputation
\systemname accepts relative to the best current cache location.
Increasing $\eta$ from 0 to 0.4 expands the mean cache-admissible set
from 4.1 to 6.7 instances while maintaining a KVC hit rate of
80.4--81.2\% (Fig.~\ref{fig:eta-sensitivity}(a)--(b)).
At $\eta=1$, the set reaches 15.9 instances, but the hit rate falls
by 7.6\,pp to 72.9\%.
Thus, a moderate budget exposes additional placements without materially
reducing locality, whereas removing the bound admits substantially more
recomputation.

\paragraph{Recomputation and GPU efficiency.}
From $\eta=0$ to 0.4, active GPU-hours decrease by 7.9\%, from 41.9
to 38.5.
Over the same range, pure-decode GPU time falls from 14.6 to 12.9
GPU-hours, while prefill-only GPU time falls from 27.1 to 25.5
GPU-hours (Fig.~\ref{fig:eta-sensitivity}(c)).
Further relaxation reverses this benefit.
Compared with $\eta=0.4$, $\eta=1$ saves only 0.4 pure-decode GPU-hours
but adds 9.6 prefill GPU-hours, increasing active GPU-hours by 23.7\%.
These results show that lost KVC reuse can erase the decode savings from
greater placement flexibility, motivating an explicit recomputation bound.
The tested budgets from $\eta=0.2$ to 0.5 use 38.5--39.0 active GPU-hours
while keeping P90 mTPOT below the 50\,ms SLO
(Fig.~\ref{fig:eta-sensitivity}(d)), providing a moderate operating range.

\subsubsection{TPOT Predictor Comparison}
\label{sec:predictor-comparison}

We compare \systemname's profile-based predictor (\systemname) with an XGBoost~\cite{xgboost2016}
variant under the 50\,ms setup of \S\ref{sec:end-to-end}, with all other strategy 
parameters remaining fixed. 
Given \systemname's higher observed accuracy, we also equip llm-d with \systemname (llm-d+) in
the end-to-end evaluation (\S\ref{sec:end-to-end}) for a fairer comparison of scheduling policies.

\begin{figure}[t]
  \centering
  \includegraphics[width=\columnwidth]{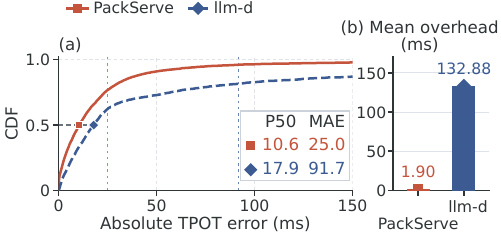}
  \caption{\systemname has lower TPOT errors (a) and mean scheduling latency (b).}
  \label{fig:predictor-diagnostics}
\end{figure}

\paragraph{Prediction accuracy.}
For the black-box baseline, we instantiate llm-d's prediction method
using its framework and train the XGBoost model on roughly 100k samples.
Fig.~\ref{fig:predictor-diagnostics}(a) shows that \systemname's absolute-error CDF
lies consistently to the left of llm-d's,
indicating smaller prediction errors across the displayed range.
\systemname achieves a P50 absolute error of 10.6\,ms, compared with
17.9\,ms for llm-d. 
The median error is roughly one-fifth of the 50-ms SLO. Together with the observed minute-level SLO control (Fig. 6(c)), 
these results support \systemnameplain's practical use in online SLO-aware scheduling.
Meanwhile, \systemname's residual error can reflect prefill interference from later
arrivals, which cannot be fully anticipated at scheduling time.
In contrast, llm-d's predictor shows larger errors, which may reflect
insufficient training coverage of runtime workload states.

\paragraph{Scheduling overhead.}
For scheduling overhead, using \systemname's predictor yields a mean
arrival-to-decision latency of 1.9\,ms, compared with 132.9\,ms using
llm-d's predictor
(Fig.~\ref{fig:predictor-diagnostics}(b)).
llm-d's separately deployed prediction service can incur further
communication costs not measured here.
\systemname's compact arithmetic therefore offers a practical advantage
for latency-sensitive online scheduling.

\begin{figure}[t]
  \centering
  \includegraphics[width=\columnwidth]{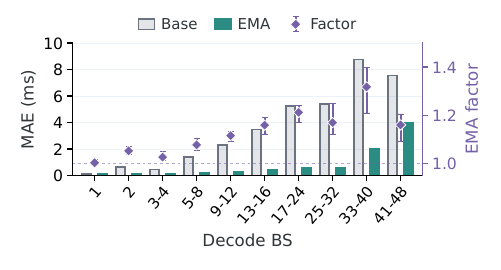}
  \caption{Pure-decode MAE (bars, left) and EMA factors (median and P10--P90, right) by BS region.}
  \label{fig:ema-tpot}
\end{figure}

\paragraph{Effect of online calibration.}
\phantomsection
\label{sec:ema-tpot}
BS-specific EMA correction addresses the profiled model's systematic
underestimation at high concurrency (\S\ref{sec:online-calibration}).
To evaluate EMA calibration, we compare predicted and measured pure-decode
iteration times in the 50\,ms run.  Unlike request-level TPOT, these
measurements exclude delays caused by prefill interference.
Across BS$\geq17$, the base predictor's mean signed error is $-5.96$\,ms,
matching its mean absolute error (MAE) in magnitude.
Applying the logged EMA factors reduces MAE from 5.96 to 0.85\,ms
(Fig.~\ref{fig:ema-tpot}).
The weighted absolute percentage error, defined as total absolute error
divided by total measured duration, falls from 17.5\% to 2.5\%.
These results support BS-specific calibration to reduce optimistic
decode estimates on packed instances.

\subsubsection{Impact of Instance Selector}
\label{sec:selector-ablation}

We compare selection priorities within \systemname's TPOT-admissible set
under the 50\,ms setup of \S\ref{sec:end-to-end}, while holding the
remaining strategy parameters constant.

\begin{table}[t]
\centering
\caption{Instance selector ablation.}
\label{tab:selector-ablation}
\fontsize{8}{9.5}\selectfont
\renewcommand{\arraystretch}{1.0}
\setlength{\tabcolsep}{3pt}
\setlength{\fboxsep}{1.2pt}
\setlength{\fboxrule}{0.3pt}
\begin{tabularx}{\columnwidth}{Xrrrr}
\toprule
Selector & GPU-h & KVC hit (\%) & P90 mTPOT (ms) & Req P90 (ms) \\
\midrule
Random & 48.09 & 77.78 & \fbox{21.61} & \textbf{63.38} \\
D & 39.97 & 78.88 & \fbox{49.80} & 89.56 \\
D--Q & 40.27 & 78.95 & \fbox{44.97} & 80.84 \\
D--C & 40.23 & 79.25 & \fbox{47.42} & 83.57 \\
\midrule
D--Q--C & 39.82 & 79.60 & \fbox{47.30} & 85.32 \\
H--C--Q & 45.73 & 76.86 & \fbox{42.82} & 129.19 \\
\rowcolor{PaperVermilion!8}
D--C--Q & \textbf{38.86} & \textbf{80.17} & \fbox{44.66} & \underline{80.25} \\
\bottomrule
\end{tabularx}
\par\vspace{2pt}
\begin{minipage}{\columnwidth}
\fontsize{8}{9.5}\selectfont\raggedright
Random: uniform. Scheduler-side priorities: D, decoding requests$\uparrow$;
C, predicted cached tokens$\uparrow$; Q, queued requests$\downarrow$;
H, predicted TPOT headroom$\downarrow$.
Bold: best GPU-h/KVC/Req P90; underline: Req P90 runner-up;
boxes: P90 mTPOT $\le50$\,ms; shading: adopted selector.
\end{minipage}
\end{table}

\paragraph{Headroom-first versus decode-BS-first selection.}
\systemname prioritizes the instance with the most scheduler-tracked
decoding requests (D), while H prioritizes the smallest predicted TPOT
headroom which is used in llm-d.
With P90 mTPOT below 50\,ms in both runs, D--C--Q uses
15.0\% fewer active GPU-hours and has 37.9\% lower request-level
P90 TPOT than H--C--Q (Table~\ref{tab:selector-ablation}). 
Small headroom can reflect long contexts or prefill interference as 
well as high concurrency. (model in \S\ref{sec:tpot-admission})
D directly concentrates decoding requests within the candidate set,
favoring larger decode batches after checking the latency constraint.

\paragraph{Cache-aware tie-breaking.}
C favors prefix reuse among instances with equal decode counts.
Moving C before Q reduces active GPU-hours by 2.4\% and improves
KVC hit by 0.6\,pp, giving D--C--Q the lowest observed footprint
and highest KVC hit rate among the tested selectors
(Table~\ref{tab:selector-ablation}).
These results support C before Q in our D--C--Q ordering
(\S\ref{sec:consolidation-overload}).

\subsection{Deployment in Production}
\label{sec:production-evaluation}

\begin{figure}[t]
  \centering
  \includegraphics[width=\columnwidth]{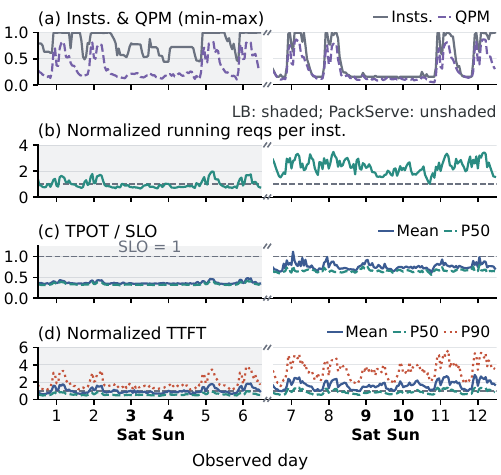}
  \caption{Resource usage and latency over 12 matched days. Breaks omit one day.}
  \label{fig:production-online}
\end{figure}

Our production deployment serves an approximately 700B-parameter model
on a cluster with over 1,000 GPUs.
We compare production cache-aware load balancing (LB) with \systemname
using two six-day windows (12 days total), one week apart and matched
by weekday and time of day (Fig.~\ref{fig:production-online}).
For confidentiality, metrics are normalized or shown without absolute
scales.

Request packing leads to a smaller serving footprint and higher
per-instance concurrency. Fig.~\ref{fig:production-online}(a) shows that the
mean instance count is 34.7\% lower than under LB, while QPM follows
similar patterns under both policies.
Meanwhile, running requests per instance rise to $2.3\times$ the LB mean and remain elevated
through most of the \systemname period (Fig.~\ref{fig:production-online}(b)).
The footprint reduction is larger on weekends than on weekdays
(54.2\% versus 26.7\%), demonstrating the effectiveness of \systemname 
under lighter workloads.
This footprint reduction is achieved while keeping P50 TPOT below the SLO.
Although higher than under LB, the P50 TPOT curve peaks at 79.8\%
of the target during the \systemname period
(Fig.~\ref{fig:production-online}(c)).
These gains do not come for free, however, as mean TTFT increases
by 66.4\% relative to LB
(Fig.~\ref{fig:production-online}(d)).
Overall, this tradeoff is acceptable in our production deployment,
supporting \systemname's effectiveness in reducing the serving footprint.

\section{Related Work and Discussion}
\label{sec:discussion}

\systemname targets colocated prefill/decode and instance-local KVC reuse.
Global KV pools and P/D disaggregation relax these assumptions while
complementing its cost-based admission and request packing.

\paragraph{Global KV Pool.}
Growing session histories, tool outputs, and external documents make
long-prefix recomputation increasingly costly.
Mooncake~\cite{mooncake2025}, MemServe~\cite{memserve2024}, and
LMCache~\cite{lmcache2025} extend cross-instance KVC reuse beyond GPU
memory to CPU memory, SSDs, and remote storage, while
Tutti~\cite{tutti2026} optimizes SSD-backed retrieval.
When retrieval is cheaper than recomputation, it can lower TTFT and
reduce prefill interference with colocated decoding.
\systemname could incorporate location- and tier-dependent lookup and
transfer costs, together with remaining computation, into an effective
prefill-cost model while retaining its SLO-admission and
instance-consolidation structure.

\paragraph{P/D Disaggregation.}
DistServe~\cite{distserve2024}, Splitwise~\cite{splitwise2023}, and
Mooncake~\cite{mooncake2025} separate prefill and decode to avoid direct
execution interference and provision resources for TTFT and TPOT
independently.
Prefill-as-a-Service (PrfaaS)~\cite{prfaas2026} further explores
selectively offloading long-context prefill across datacenters and
returning the resulting KVC to local decode pools.
Fully separated execution would remove the direct P/D interference
term from our TPOT model, reducing one source of prediction uncertainty
and potentially simplifying SLO-based isolation.
Extending \systemname would still require modeling KVC transfer,
network variability, and queues on both sides, as well as jointly
selecting prefill and decode destinations.
We leave these extensions to future work.

\section{Conclusion}
\label{sec:conclusion}

This paper presents \systemname, a gateway-level request scheduler designed
to preserve KVC reuse, meet workload-specific TPOT SLOs, and reduce the GPU
footprint of agentic LLM serving. Motivated by our characterization of
production workloads, \systemname balances the throughput gains from request
packing against the prefill cost of lost cache reuse. We develop compact
white-box models for accurate, low-overhead latency prediction under
prefill/decode interference. Guided by these models, \systemname prioritizes
KVC reuse by bounding additional prefill recomputation, then favors instances
with higher decode concurrency among those predicted to meet the TPOT target.
Experiments on 64 NVIDIA H20 GPUs show that \systemname uses 13.0--24.6\% fewer active GPU-hours than
LMetric and llm-d+ while meeting the evaluated windowed TPOT objectives.
A 12-day study across a production cluster with over 1k GPUs further shows 34.7\% fewer
serving instances and 36.8\% higher per-instance request throughput relative
to the production cache-aware load-balancing policy.

\par

\bibliographystyle{ACM-Reference-Format}
\bibliography{references}
\end{document}